\documentclass[aip,a4paper,preprintnumbers,groupedaddress,showkeys
]{revtex4-1} 
\usepackage{overpic}
\usepackage{ulem}
\usepackage{epstopdf,epsfig}
\usepackage{tikz}
\usepackage[fleqn]{amsmath}
\usepackage{amssymb}
\usepackage{bm}
\usepackage{graphicx} 
\usepackage{subfigure}
\usepackage{booktabs}
\usepackage{mathrsfs}
\usepackage{color}
\usepackage{setspace}
\usepackage{upgreek}
\usepackage{tabularx}
\usepackage{appendix}
\usepackage{CJK}
\usepackage[pdfstartview=FitH,
            CJKbookmarks=true,
            bookmarksnumbered=true,
            bookmarksopen=true,
            colorlinks=true,
            pdfborder=001,
            linkcolor=blue,
            citecolor=blue,
            urlcolor=blue
            ]{hyperref}
\RequirePackage{color}
\begin{document}
\begin{CJK*}{UTF8}{gbsn}
	
\title{Numerical simulation of flow field and debris migration in extreme ultraviolet source vessel}

\author{Wen-Sheng Meng (孟文盛)$^{1}$}
\altaffiliation{These authors contributed equally to this work}
\author{Chao-Ben Zhao(赵超本)$^{1}$}
\altaffiliation{These authors contributed equally to this work}
\author{Jian-Zhao Wu (吴建钊)$^{1}$}
\author{Bo-Fu Wang (王伯福)$^{1}$}
\email[Authors to whom correspondence should be addressed: ]{bofuwang@shu.edu.cn and qzhou@shu.edu.cn and klchong@shu.edu.cn}
\author{Quan Zhou (周全)$^{1}$}
\email[Authors to whom correspondence should be addressed: ]{bofuwang@shu.edu.cn and qzhou@shu.edu.cn and klchong@shu.edu.cn}
\author{Kai Leong Chong (庄启亮)$^{1}$}
\email[Authors to whom correspondence should be addressed: ]{bofuwang@shu.edu.cn and qzhou@shu.edu.cn and klchong@shu.edu.cn}
\affiliation{$^1$ Shanghai Key Laboratory of Mechanics in Energy Engineering, Shanghai Institute of Applied Mathematics and Mechanics, School of Mechanics and Engineering Science, Shanghai University, Shanghai, 200072, China}

\date{\today}

\begin{abstract}
	
Practical extreme ultraviolet (EUV) sources yield the desired 13.5 nm radiation but also generate debris, significantly limiting the lifespan of the collector mirror in lithography. In this study, we explore the role of buffer gas in transporting debris particles within a EUV source vessel using direct numerical simulations (DNS). Our study involves a 2m $\times$ 1m $\times$ 1m rectangular cavity with an injecting jet flow subjected to sideward outlet. Debris particles are introduced into the cavity with specified initial velocities, simulating a spherical radiating pattern with particle diameters ranging from 0.1 $\upmu$m to 1 $\upmu$m. Varying the inflow velocity (from $1$m/s to $50$m/s) of the buffer gas reveals a morphological transition in the flow field. At low inflow velocities, the flow remains steady, whereas higher inflow velocities induce the formation of clustered corner rolls. Upon reaching sufficiently high inflow velocities, the jet flow can penetrate the entire cavity, impacting the endwall. Interestingly, the resulting recirculation flow leads to the spontaneous formation of spiraling outflow. The distinct flow structures at various inflow velocities lead to distinct patterns of particle transport. For low-speed gas, it is efficient in expelling all particles smaller than 0.4 $\upmu$m, while for high-speed gas, those fine particles accumulate near the endwall and are challenging to be extracted. Our findings highlight the significance of controlling flow conditions for effective debris particle transport and clearance in diverse applications especially in EUV source vessels.

\end{abstract}

\maketitle
\end{CJK*}

\section{Introduction} \label{sec_intro}

The state-of-the-art technology for producing microchips relies on the utilization of extreme ultraviolet (EUV) lithography, which has emerged as a promising approach \cite{Johnston_pof_1973, Gerry_sp_2005,Pirati_sp_2017, Hutcheson_sp_2018,bakshi2018euv,fu2019euv}. Among the various methods of generating EUV light, the Laser-produced plasma (LPP) scheme utilizing carbon dioxide ($\mathrm{CO}_2$) laser irradiation on tin (Sn) droplets has gained significant attention \cite{Mainfray_rpp_1991, Akira_sp_2007, Aota_jpcs_2008, Nowak_oer_2013, Mark_sp_2020}. This approach generates light with a wavelength of $13.5$ nm, making it a feasible candidate for EUV lithography in semiconductor manufacturing.

\begin{figure}[th]
	\includegraphics[width=1.0\linewidth]{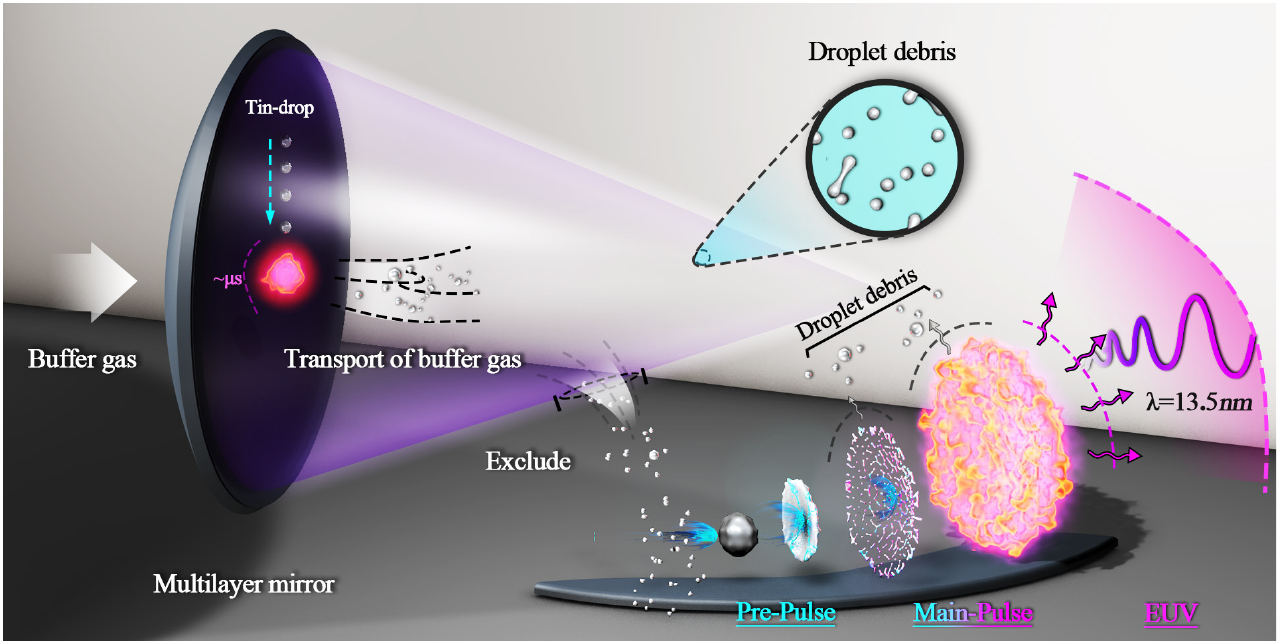}
	\caption{The process of Sn droplet debris formation in LPP. A high-intensity laser rapidly vaporizes a tin (Sn) droplet, creating an energetic plasma. This plasma then quickly expands due to pressure and temperature gradients, and undergoes cooling and recombination, leading to the formation of solid or liquid Sn particles. These particles can be transported by the buffer gas flow or deposited nearby, analyzed as particle-laden flows.}
	\label{fig:fig1}
\end{figure}

LPP utilizes high-intensity lasers to generate highly ionized gas or plasma, spanning different time and length scales \cite{Johnston_pof_1973, Mainfray_rpp_1991, David_sp_2009, Poirier_jqsrt_2006}, which presents multi-scale and multi-physics features. While the primary objective of LPP is to generate EUV light, the formation of debris, specifically during the interaction between the laser and Sn droplets, remains a by-product of this process \cite{Aota_jpcs_2008, Ando_apl_2006, Harilal_ol_2006, Banine_jpd_2011}. The general physical steps encompassing LPP include plasma formation, plasma expansion, plasma cooling, recombination, as well as Sn debris formation as shown in Fig.~\ref{fig:fig1}. The details of each process are as follows:

\begin{itemize}
    \item Plasma Formation: The Sn droplet is subjected to a focused high-intensity laser beam, resulting in rapid localized vaporization and the formation of a highly energetic plasma. Before the illumination by the main laser pulse, there is a pre-pulse in advance to enhance the efficiency of generating EUV light.
    \item Plasma Expansion: The plasma undergoes rapid outward expansion, driven by high pressure and temperature gradients arising from plasma formation. This process necessitates the analysis of compressible flow, occurring within an extremely short timescale (on the order of nanoseconds).
    \item Plasma Cooling and Recombination: Adiabatic expansion causes the plasma to cool down as it interacts with the surrounding environment. Cooling leads to the recombination of ions and electrons, resulting in the formation of neutral species.
    \item Sn Debris Formation: During plasma expansion and cooling, solidification or condensation of Sn occurs, giving rise to the formation of small solid or liquid particles. These debris particles can either be transported away from the target region by the gas flow or deposited on nearby surfaces. The transport and deposition of debris particles can be analyzed within the framework of particle-laden flows.
\end{itemize}

The deposition and transport of debris particles are significant concerns in EUV lithography. The deposition of tin (Sn) debris causes contamination and decreases the reflectivity of the collector mirrors. As the molten tin accumulate on the surface, 1 nm of tin can lead to the 10\% degradation of the mirror reflectivity \cite{bakshi2018euv}. To address this, various methods have been explored by different studies. These methods include the utilization of cavity-confined plasma, tape targets \cite{Steven_ao_1993}, mass-limited droplet targets \cite{Richardson_jvs_2004,Takenoshita_ps_2005}, foil trap\cite{Joseph_ps_2005}, electrostatic repeller fields\cite{Chiew_ps_2005}, and application of magnetic fields\cite{Harilal_jap_2007,Gohta_ps_2003}. So far, a common approach relies on the use of buffer gas with high flow rate in addressing debris-related challenges in EUV lithography \cite{Shmaenok_ps_1998,Bollanti_apb_2003,Klunder_ps_2005,Wouter_ps_2006,Harilal_apb_2007,Bleiner_jap_2009,Abramenko_apl_2018}. Bollanti et al. \cite{Bollanti_apb_2003} employed krypton as ambient gas to control debris generated from laser-produced Ta plasma, resulting in a significant reduction in debris accumulation at the plate. However, careful selection of the ambient environment is crucial for EUV lithography, given that many gases tend to absorb EUV light. Hydrogen, helium, and argon have been identified as having exceptional transmission capability at 13.5 nm\cite{Harilal_apb_2007}. Wouter et al. \cite{Wouter_ps_2006} explored directional gas flows as a strategy to mitigate debris in practical EUV lithography. Their experiments demonstrated effective suppression of microparticles through transversal gas flow, presenting potential applications in combination with a foil trap. In the meantime, Abramenko et al.\cite{Abramenko_apl_2018} investigated the mitigation of fast ions from laser-produced Sn plasma using background gas to protect EUV collector optics in lithography. Importantly, the presence of buffer gas at a specific pressure did not affect the transmission of light. Laser-induced fluorescence imaging visualized the sputtering of a dummy mirror, with gas reducing the sputtering rate by approximately $\sim$5\%. Among these approaches, the use of buffer gas stands out as particularly noteworthy for mitigating debris-related issues. These studies collectively underscore the significance of introducing buffer gas into the chamber to flush out particles, thereby maintaining the cleanliness of optical components in EUV lithography.

In the present study, we analyze the effect of injected buffer gas on the overall flow structure within the EUV source vessel. We further examine how varying inlet velocities of buffer gas influences particle transport. Our objective is to uncover the mechanisms governing particle transport when buffer gas is present in a lithographic setting. The remainder of the paper is organized as follows. The physical model and numerical setup are briefly introduced in Sec.~\ref{Numerical setups}. Then, the results are shown in Sec.~\ref{Result and discussion}, we focus on the influence of buffer gas on the overall flow patterns in Sec.~\ref{Global features} and examine the variations in debris particle trajectories in Sec.~\ref{Debris particle trajectory}. Concluding remarks are given in Sec.~\ref{Conclusion}.

\section{Numerical Setups}\label{Numerical setups}
\subsection{Flow simulation}

\begin{figure}[th]
	\includegraphics[width=1.0\linewidth]{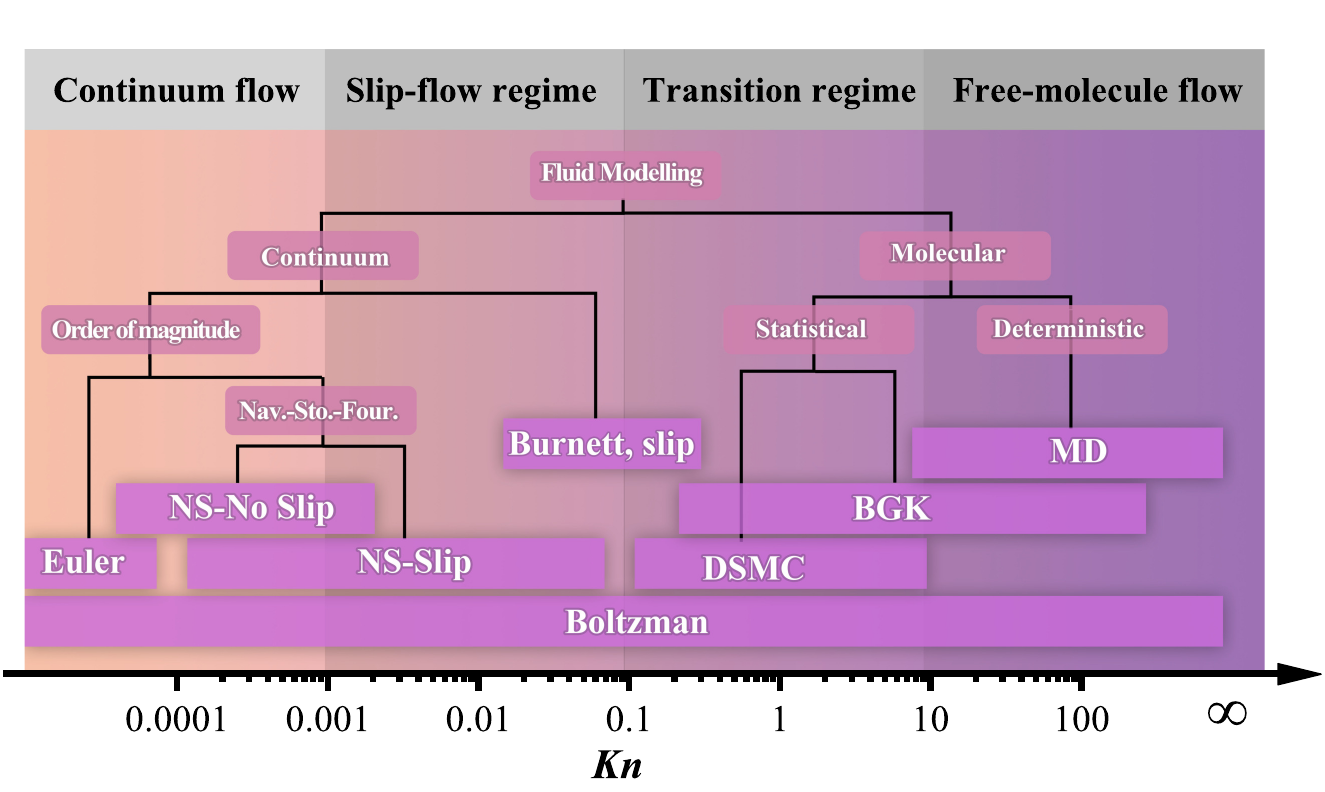}
	\caption{Classification of flow regimes and models based on the work of Gad-el-Hak\cite{Gad_jfe_1999}. The practical usability of gas flow models is depicted as a function of the Knudsen number $Kn$. The limits used in this classification are taken from various references\cite{Gad_jfe_1999,Struchtrup_ima_2011,Veltzke_pof_2012,song_ijhmt_2015,Mouro_jmm_2021,Alafnan_pof_2022}. It is anticipated that with advancements in computational power, the lower limits of numerical methods such as direct simulation Monte Carlo (DSMC) and molecular dynamics (MD) will be shifted to higher Knudsen numbers in the future.}
	\label{fig:kn}
\end{figure}

We focus on the debris migration in EUV source vessels, characterized by relatively low ambient pressure. In such circumstances, it becomes crucial to initially evaluate the rarefaction effects within the fluid flow. Rarefaction effects are typically characterized by the Knudsen number, which is defined as the ratio between the molecular mean free path of a gas and a characteristic dimension of the flow domain. The definition of Knudsen number is
\begin{eqnarray}
Kn = \frac{\overline{\lambda}}{\widetilde{L}},\label{eq:01}
\end{eqnarray}
where $\widetilde{L}$ is the characteristic length of the vessel and $\overline{\lambda}$ is the molecular mean free path. The value of $\overline{\lambda}$ can be obtained as follows\cite{Alafnan_pof_2022}:
\begin{eqnarray}
 \overline{\lambda} =\frac{k_B T}{\sqrt{2}\pi d^2 P},\label{eq:02}
\end{eqnarray}
where $k_B$ is the Boltzmann constant, $T$ is the temperature, $d$ is the molecular diameter, and $P$ is the pressure. The Knudsen number is used to demarcate the fluid flow into four regimes \cite{Gad_jfe_1999,Struchtrup_ima_2011,Veltzke_pof_2012,song_ijhmt_2015,Mouro_jmm_2021,Alafnan_pof_2022} as shown in Fig. \ref{fig:kn}: continuum regime $Kn < 10^{-3}$, slip regime $10^{-3}< Kn <10^{-1}$, transition regime $10^{-1}< Kn <10$, and free molecular regime $Kn >10$. Note that these regimes can even co-exist in a given passage. It can be seen that the continuum flow models are capable of describing the transport for $Kn$ values less than 0.1, while $Kn$ greater than 10 is deemed to be governed by free molecular diffusion. In our study, we adopt hydrogen (${H_2}$) as buffer gas and consider its working pressure of $P$ = 100Pa in EUV source vessel. Based on Eq.\ref{eq:02}, by taking the height of the cavity as characteristic length scales, we can estimate that $Kn$ $\approx$ 2.29$\times 10^{-4}$, which is less than 1.0$\times 10^{-3}$. Therefore, the flow in the domain can be modeled using continuum fluid mechanics (Navier-Stokes equations with no-slip boundary conditions). A detailed list of parameters for hydrogen and debris particles can be found in Table \ref{tab:tab1}.

\begin{table}[th]
	\centering 
	\caption{List of parameters for hydrogen and debris particle. These parameters are also employed in the numerical simulations in this study.}
	\label{tab:tab1}
	
	\begin{ruledtabular}
		\begin{tabular}{cccccc}
			Properties &Symbols &Values&Properties &Symbols &Values\\
			\hline \\
			Temperature& $T$(K) &295&Boltzmann constant& $k_B$ (JK$^{-1}$) &  1.38$\times 10^{-23}$ \\
			Characteristic length& $\tilde{L}$ (m) &  1& Molecular diameter & $d$ (m) &$ 2\times 10^{-10}$ \\
			Pressure& $P$ (Pa) &  100&Gas dynamic viscosity & $\mu$(Pa s$^{-1}$)& 8.92$\times 10^{-6}$\\
			Sutherland constant&$S_{H_2}$ (K)& 97& Partical density(Sn)
			& $\rho_p$(kg m$^{-3}$)& 6.97$\times 10^{4}$\\
			Particle diameter & $d_p$ ($\upmu$m) & 0.1$\sim$1.0&Gas density(H$_2$)& $\rho$ (kg m$^{-3}$) & 8.96$\times 10^{-5}$ 
		\end{tabular}
	\end{ruledtabular}
\end{table}
The directional gas flow is injected to flush the particles. The governing equations of the gas flow (justified to be within the continuum regime) are given by
\begin{eqnarray} 
    &\nabla \cdot \boldsymbol{u} = 0,\label{eq:03}\\
    & \displaystyle \frac{\partial \bm{u}}{\partial t} + \bm{u}\cdot \nabla \bm{u} = -\frac{1}{\rho} \nabla p + \nu \nabla^2 \bm{u},\label{eq:04}
\end{eqnarray}
Here, $\bm{u}$ is the gas velocity, $\rho$ the gas density, $p$ the pressure, and $\nu$ the kinetic viscosity. To simulate the gas flow, the governing equations are numerically solved by a second-order finite difference
code, which has been widely validated in the literature \cite{Verzicco_jcp_1996,Ostilla_jcp_2015,Poel_cf_2015}. The third-order Runge-Kutta method combined with the second-order Crank-Nicholson scheme is applied for time integration. More details on the numerical approach can be found in previous studies\cite{ostilla2015multiple,Chong_jcp_2018,guo_jfm_2023}.

\subsection{Debris simulation}
Different from simulating the gas flow by the Eulerian approach,  the motion of debris particles is simulated by employing the Lagrangian method. Here, we focus on only the latter stage of debris formation where small solid or liquid particles form after cooling down, and the debris particle is assumed to be electrically neutral. We employ the spherical point-particle model, taking into account the momentum equation as follows  \cite{Maxey_pof_1983, Gatignol_jmta_1983, Tsai_taml_2022} 

\begin{eqnarray} 
\frac{du_{i,n}}{dt}= (\beta+1)\frac{Du_{i,g,n}}{Dt}+(\beta+1)\frac{12\nu(u_{i,g,n}-u_{i,n})}{d^2_{p}}f_d+g \beta \hat{e}_z \label{eq:05}. 
\end{eqnarray}
 where $u_{i,n}$ and $u_{i,g,n}$ are the velocities of particles and gas at the location of the particles, respectively. $d_{p}$ is the particle diameter. $\beta$ is a dimensionless measure of the particle density relative to the buffer gas density and is defined as $\beta \equiv 3\rho/(\rho+2\rho_p)-1$ with $\rho_p$ the density of particles. In this work, we consider spherical particles with a density much larger than that of fluid, i.e., $\rho_p  \gg \rho$. For these heavy particles, the dominant forces are the drag force and gravity with negligibly small effects from other forces. It is worth noting that the particle size is quite small, one should take into account the rarefaction effects sensed by the particles. As a result, the drag coefficient $f_d$ is modified and expressed as a function of the particle Reynolds number $Re_p$ and the particle Knudsen number $Kn_p$:

 \begin{eqnarray}
 f_d=\frac{1+0.169Re^{2/3}_{p}}{S(Kn_p)} \label{eq:08},
 \end{eqnarray}
 where $Re_p = \rvert u_{i,n} -u_{i,g,n}\rvert d_p /\nu$, $Kn_p = \overline{\lambda} / d_p$. The particle Knudsen number $Kn_p$ is defined using the particle diameter as the characteristic length scale. Here, $Kn_p$ $\approx$ 4$\times 10^{4} \gg 1$, characterizes the flow as a series of discrete ballistic collisions between gas molecules and the particle \cite{Peter_ast_2004}. From the particle perspective, the gas flow is quite rarefied, and a correction to the drag equation becomes crucial. As outlined by Hinds et al. \cite{hinds_aerosol_2022}, the implementation of this correction involves the use of the Cunningham slip correction factor, $S(Kn_p)$, parameterized by Allen \& Raabe \cite{Allen_jas_1982, Michael_ast_1985} as
 \begin{eqnarray}
S(Kn_p) = 1 + Kn_p \left[ C_1 +  C_2 \mathrm{exp} \left( \frac{-C_3}{Kn_p}\right)  \right] \label{eq:010}.
 \end{eqnarray}
 
 Here $C_1$=2.514, $C_2$= 0.8, and $C_3$= 0.55 are empirically determined constants specific to the system under consideration. Notably, these values are subject to variation depending on the suspending gas, especially when it differs from the air at normal temperature (298K) and standard atmospheric pressure, as indicated by Rader\cite{Brockmann_ast_1990}. Furthermore, it is essential to highlight that the values of $C_1$, $C_2$, and $C_3$ will vary for different equivalent diameters.

In the debris simulation, the particle velocities are interpolated from the instantaneous Eulerian velocity field using a tri-cubic polynomial interpolation scheme. This approach for simulating fine particles or droplets have been adopted in our previous studies\cite{chong2021extended,ng2021growth,zhao2023human}.

\subsection{Numerical set-up}
\begin{figure}[th]
	\includegraphics[width=0.7\linewidth]{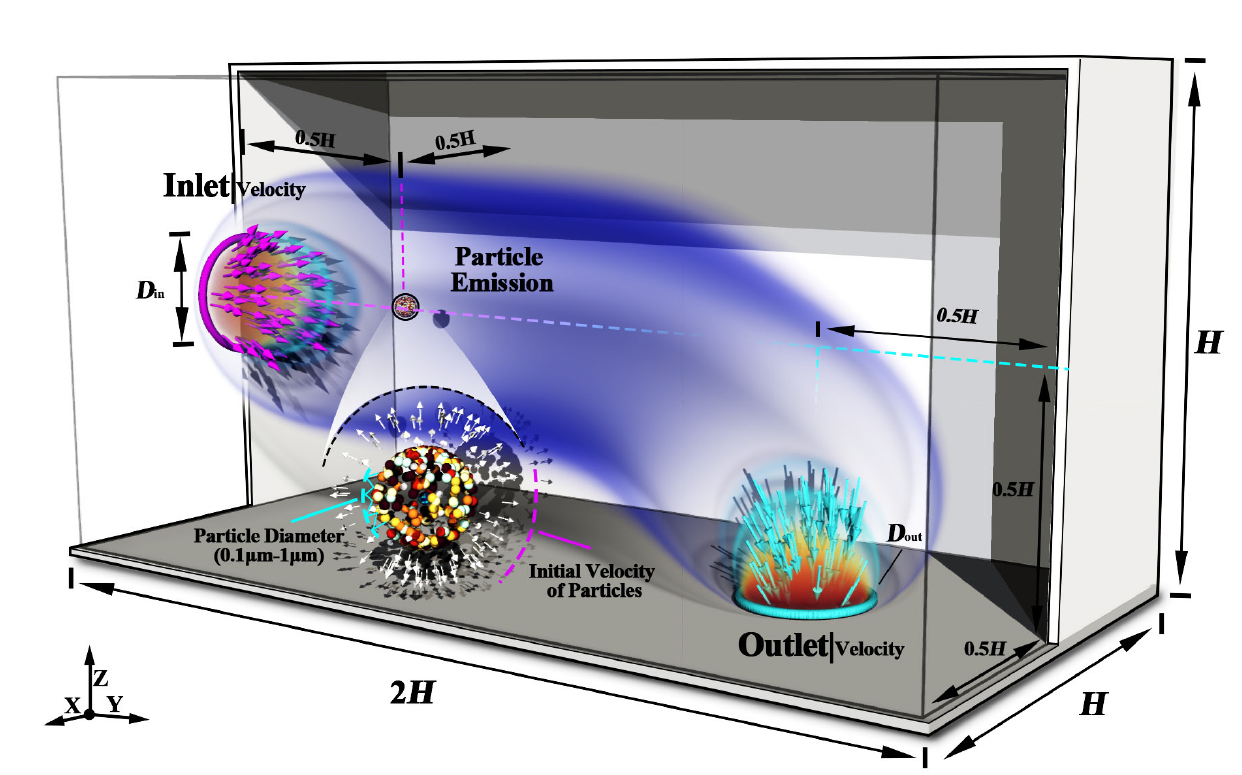}
	\caption{Illustration of the simulation setup for tin debris migration with buffer gas. Schematic details of the boundary conditions at the source used for the numerical simulations.}
	\label{fig:sketch}
\end{figure}

To investigate the debris migration in the EUV collector vessel in lithography, we consider a rectangular box with dimensions of $L:W:H = 2\ \mathrm{m} \times 1\ \mathrm{m} \times 1\ \mathrm{m}$, as illustrated in Fig.~\ref{fig:sketch}. The inlet flow is positioned at the center of the left sidewall with a diameter of $D_{\mathrm{in}}=0.25H$. On the bottom wall, the outlet is situated at the coordinates ($1.5H$, $0.5H$, 0), and it shares the same diameter as the inlet ($D_{\mathrm{in}}$=$D_{\mathrm{out}}$). The release position of the debris particles within the box has been fixed at ($0.5H$, $0.5H$, $0.5H$). The initial speed for the released particles is set at 200m/s, following a spherical radiating pattern, and the particle diameters fall within the range of 0.1 $\upmu$m to 1 $\upmu$m\cite{vinokhodov2016formation}. For each particle size, five thousand Lagrangian particles are released at their initial positions within the domain when the flow reaches a statistically stationary state.

For boundary conditions, we apply velocity boundary conditions at the inlet and outlet to ensure equal mass flow rates, while the remaining surfaces are treated as solid walls with no-slip boundary conditions. We employ a grid resolution of 128 $\times$ 64 $\times$ 64 to adequately resolve both the gradient of the jet flow.  The time step adopted is small enough to guarantee numerical stability and resolve the smooth evolution of the flow. For the sake of simplification, we opted not to couple the particles to the momentum field, considering that such coupling would have a negligible effect on the trace amount of debris. Through applying one-way coupling simulations, the continuous-phase flow field is solved and then the particle trajectory for the discrete phase is determined. In all the simulations, the flow is assumed to be incompressible with constant density and viscosity, which is reasonable for not too strong inlet. The walls were adiabatic and no-slip wall conditions were used. The simulations were performed for at least 1500s (25min), and we sampled the last 1000s (16.7min) for our statistical analysis. To ensure that the statistical steady state is reached, we compared the last 500s (8.3min) to the full sampling duration, and the variation is 1\% difference.
 
\begin{figure}[th]
	\centering\includegraphics[width=0.85\linewidth]{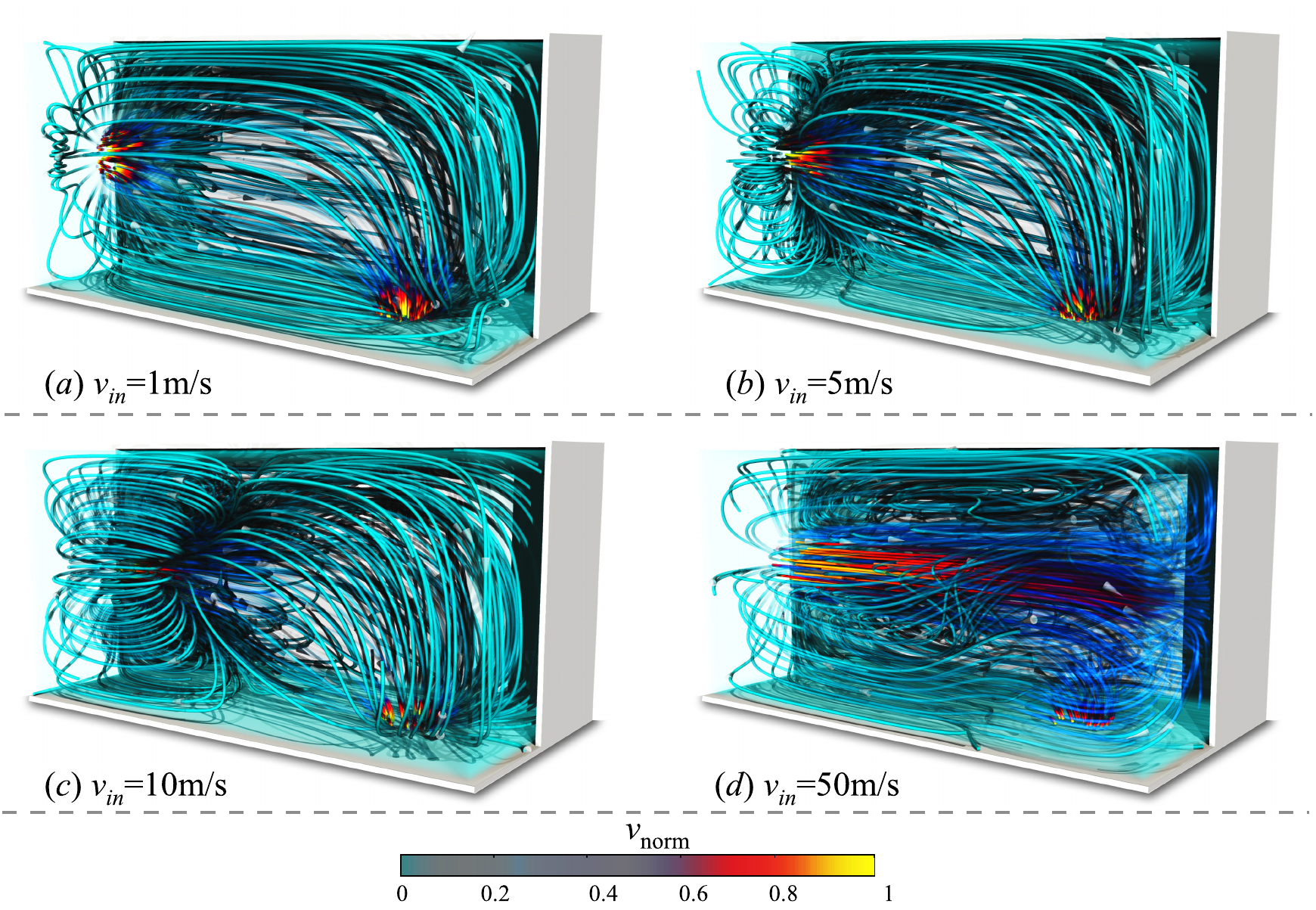}
	\caption{The instantaneous flow fields (streamlines) in the steady state for the different $v_{\rm{in}}$: $v_{\rm{in}}$=1m/s in ($a$), $v_{\rm{in}}$=5m/s in ($b$), $v_{\rm{in}}$=10m/s in ($c$), $v_{\rm{in}}$=50m/s in ($e$). The streamline color indicates the magnitude of velocity $v_{\rm{norm}}$, $v_{\rm{norm}}$=$v/v_{\rm{in}}$.}
	\label{fig:streamlines}
\end{figure}

\section{Results and discussion}\label{Result and discussion}
\subsection{Global flow features}\label{Global features}

We first examine the evolution of global flow features with increasing inflow velocity $v_{\rm{in}}$. In Fig. \ref{fig:streamlines}, we show the typical streamline of the instantaneous flow fields at different $v_{\rm{in}}$, with streamlines color-coded according to the flow speed. For a small inflow velocity, $v_{\rm{in}}$=1m/s, as shown in Fig. \ref{fig:streamlines}($a$), the flow is characterized by the initiation of streamlines at the left inlet and their termination at the bottom outlet, and the flow is steady. Notably, the strongest flow takes place in the vicinity of the inlet and outlet regions. As the inflow velocity $v_{\rm{in}}$ increases, the buffer gas jet exerts a substantial influence on the flow field, leading to the formation of clustered corner rolls at the inlet region as shown in Fig. \ref{fig:streamlines}($b$). A further increase in the inflow velocity, specifically at $v_{\rm{in}}=10$m/s, the corner rolls become lengthened as illustrated in Fig. \ref{fig:streamlines}($c$). Furthermore, at even higher inflow velocities ($v_{\rm{in}}=50$m/s), the jet is observed to penetrate through the entire cavity, impacting the right side wall and leading to the formation of a recirculation zone in the upper part of the cavity, as shown in Fig. \ref{fig:streamlines}($d$). One also observes that there is spiral motion in the vicinity of the outlet region at this extreme speed of the buffer gas.

\begin{figure}[th]
\centering\includegraphics[width=1.0\linewidth]{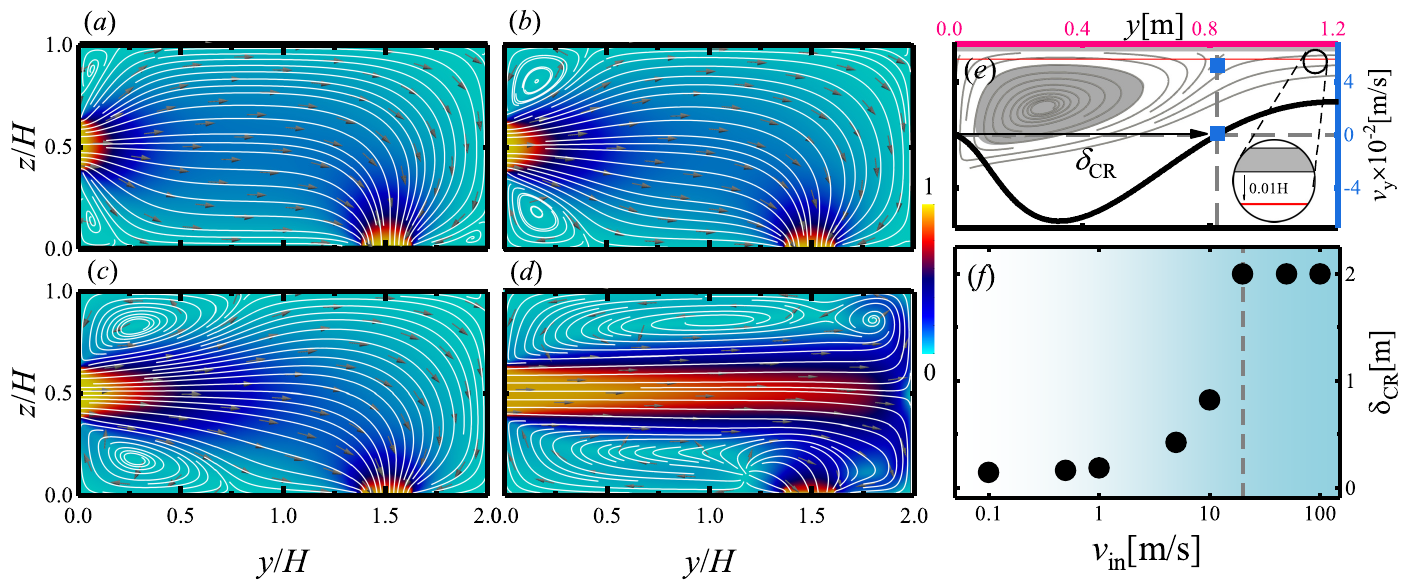}
\caption{($a$-$d$) The slices of the instantaneous velocity field with superimposed velocity vectors: $v_{\rm{in}}$=1m/s in $(a)$, $v_{\rm{in}}$=5m/s in $(b)$, $v_{\rm{in}}$=10m/s in $(c)$$v_{\rm{in}}$=50m/s in $(d)$.  The snapshots were obtained from the slice of the velocity field at mid-width. The color indicates the magnitude of velocity.  ($e$) Definition of corner roll size. The blue square marks the zero-crossing point for the $Re$= 10 case and illustrates how  $\delta_{\rm{CR}}$ is determined. Inset shows the streamlines of the velocity field at the top-left corner of the cell for the $v_{\rm{in}}$ = 10m/s case. The position of the red dashed line corresponds to the location for velocity extraction, situated at a distance of $0.01H$ from the wall. ($f$) The size of the corner rolls $\delta_{\rm{CR}}$ as a function of the $v_{\rm{in}}$. }
\label{fig:2D_streamline}
\end{figure}

To clearly demonstrate the influence of the buffer gas jet on the overall flow, we present the flow field at the mid-plane (plane $A_{\rm{yoz}}|{x=0.5W}$) with a color representation indicating the velocity magnitude in Figs. \ref{fig:2D_streamline}($a$)-($d$). From the figures, it can be observed that the dominant region of the jet increases with the inflow velocity $v_{\rm{in} }$. Additionally, the size of the corner rolls increases with the inflow velocity $v_{\rm{in}}$. To quantify the growth of the corner roll, we define the size of the top-left corner-flow roll (see Fig. \ref{fig:2D_streamline}$e$) using the horizontal distance from the sidewall to the zero-crossing or stagnation point of the horizontal velocity near the top plate at the edge of 0.01$H$. Similar quantification method has been employed in the community of studying corner roll in Rayleigh-B\'enard convection\cite{Shishkina_2014_pre,zhao_jfm_2022}. The size of the corner rolls, denoted as $\delta_{\rm{CR}}$, is evaluated and shown in Fig. \ref{fig:2D_streamline}($f$). For small $v_{\rm{in}}$, the corner roll grows monotonically with $v_{\rm{in}}$, while for sufficiently high $v_{\rm{in}}$, the value of $\delta_{\rm{CR}}$ is close to $2L$ as limited the size of the domain. This indicates that the recirculation caused by the corner roll has changed from a localized to a global flow.

\begin{figure}[h]
	\centering\includegraphics[width=0.95\linewidth]{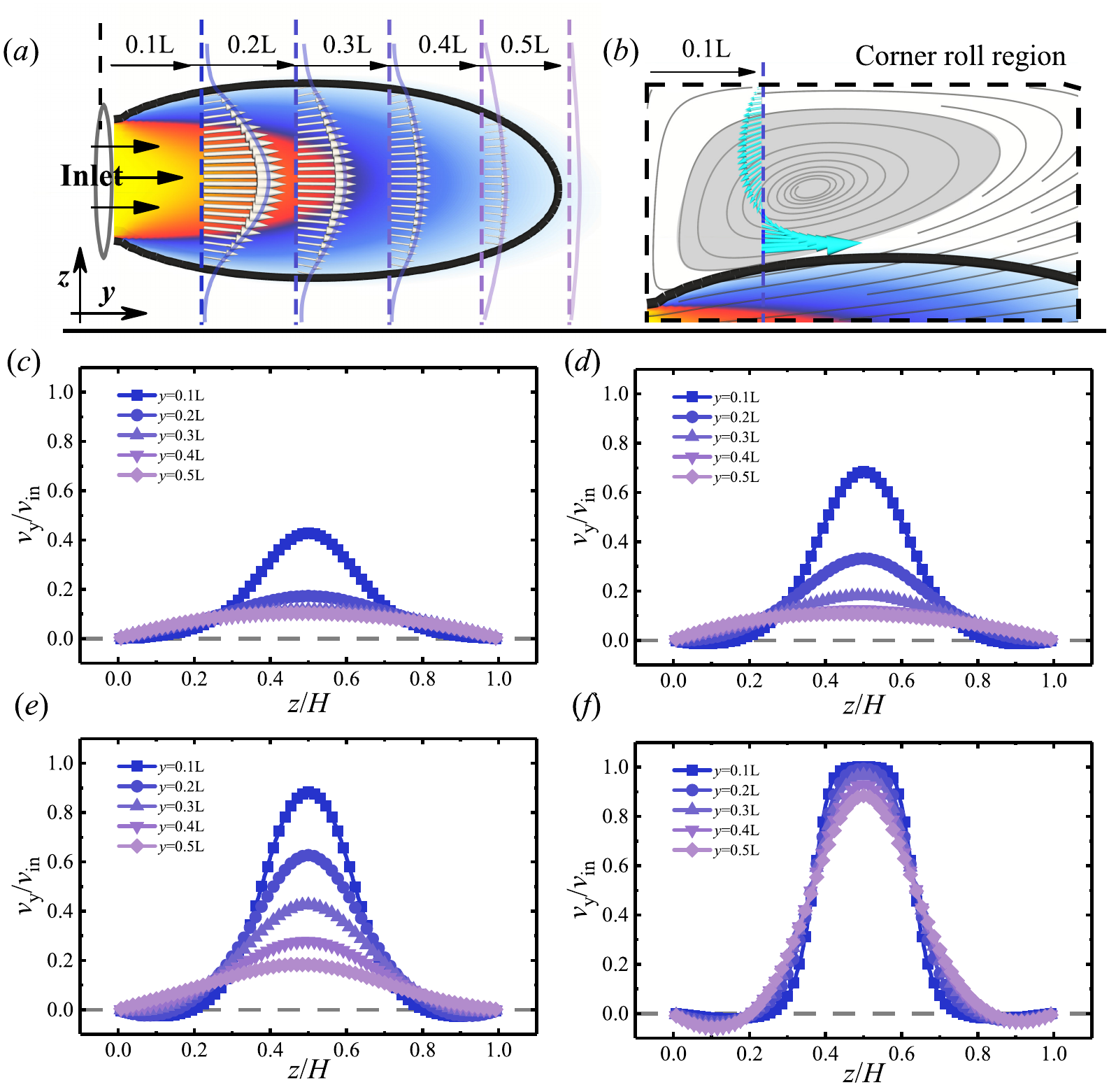}
	\caption{($a$) Evolution of velocity profiles on the vertical plane at the body mid-width. ($b$) The streamline diagram of the corner roll region with magnification, overlaid with velocity vectors at the locations where velocity profiles were extracted.  ($c$)-($f$)  The velocity ($v_y/v_{\rm{in}}$) profiles evolution downstream the patch in sections $y/H$= (0.1 0.2, 0.3, 0.4, 0.5) (as marked by the lines on the top map) for the different $v_{\rm{in}}$: $v_{\rm{in}}$=1m/s in ($c$), $v_{\rm{in}}$=5m/s in ($d$), $v_{\rm{in}}$=10m/s in ($e$), $v_{\rm{in}}$=50m/s in ($f$). }
	\label{fig:Velocity_profile}
\end{figure}

To deepen our understanding of the jet flow propagation in the cavity, we plot in Fig. \ref{fig:Velocity_profile} the evolution of velocity profiles at different longitudinal location ($y/L$=0.1, 0.2, 0.3, 0.4 and 0.5. see Fig. \ref{fig:Velocity_profile}$a$) for various inflow velocities. At a given $y/L$, the normalized profile $v_y/v_{\rm{in}}$ serves as a footprint of the jet flow, highlighting the maximum value of $v_y/v_{\rm{in}}$ at the centerline of the jet, where its magnitude diminishes longitudinally (increasing $y/L$). This dependence is due to the gradual loss of momentum of the jet to the surrounding medium. As the inflow velocity $v_{\rm{in}}$ increases further, particularly for the highest explored inflow velocities ($v_{\rm{in}}$=50 m/s in Fig. \ref{fig:Velocity_profile}$f$), smear out of the magnitude of the peak velocity becomes less distinguishable at different longitudinal locations, while there is only a slight increase in the width of the jet. It signifies that the inlet jet can remain energetic throughout its propagation in the cavity, resulting in the strong impact at the end wall. Additionally, it is noteworthy that for high inflow velocities, negative velocities can be observed near the upper and lower walls. This occurrence is attributed to localized backflow induced by the region of corner rolls (see Fig. \ref{fig:Velocity_profile}$b$).

\begin{figure}[th]
	\centering\includegraphics[width=1.0\linewidth]{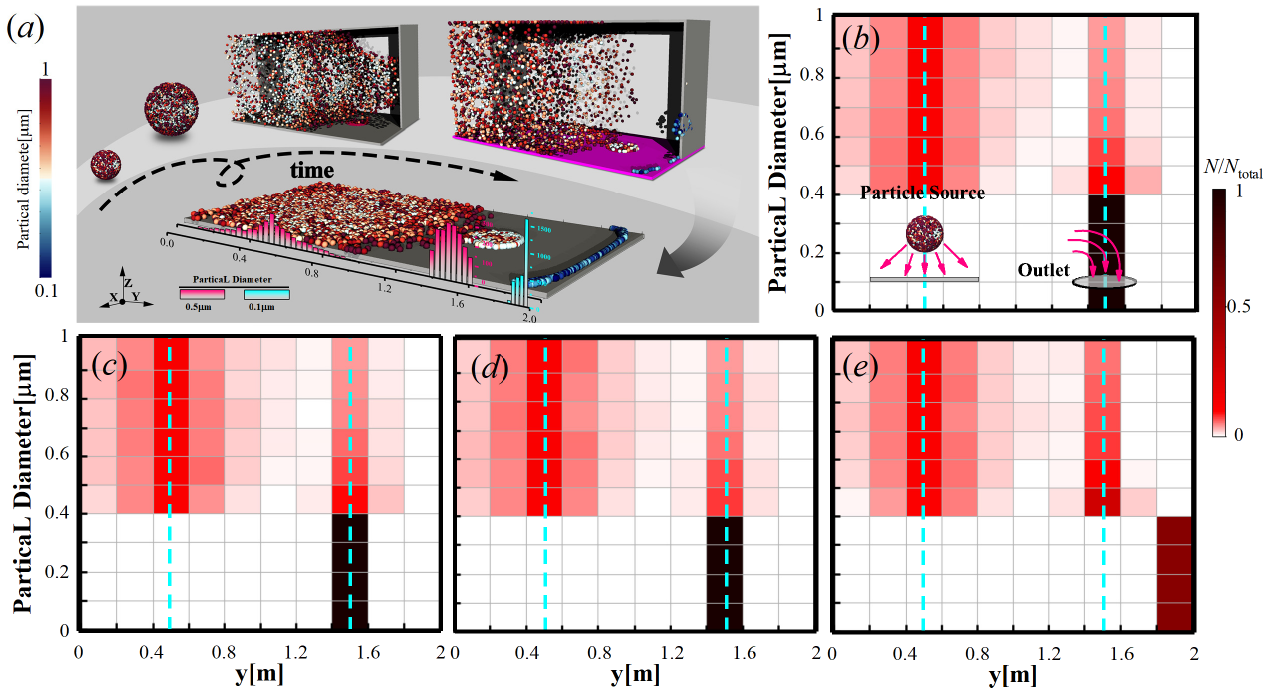}
	\caption{($a$) Flow visualization snapshots from our direct numerical simulations of debris particles at $v_{\rm{in}}$=50m/s.  Particles are color-coded by their sizes. ($b$)-($e$)The particle size histograms vs distance is presented in the plane, with contours indicating the relative values of  $N/N_{\rm{total}}$, where $N_{\rm{total}}$ represents the total number of released particles corresponding to a specific particle diameter. Panels ($b$), ($c$), ($d$), and ($e$) respectively represent the statistical results for different inflow velocities: ($b$) $v_{\rm{in}}=1$m/s, ($c$) $v_{\rm{in}}=5$m/s, ($d$) $v_{\rm{in}}=10$m/s, and ($e$) $v_{\rm{in}}=50$m/s. All histograms are based on statistical analysis of the particle field at $t$ = 500 seconds following particle release.}
	\label{fig:partical_1}
\end{figure}

\subsection{Debris particle trajectory}\label{Debris particle trajectory}
We have illustrated the emergence of corner rolls and a spiral flow structure for a sufficiently strong injection of buffer gas. It is highly desirable to understand how the change in flow structure affects the transport property of the debris particles. Recall that, for each particle size, thousands of Lagrangian particles are released at the initial positions within the domain to gain enough particle statistics. Given that the particles are released with an initial velocity of 200m/s with a spherical radiation pattern, the initial rapid expansion is evident, as shown in Fig. \ref{fig:partical_1}($a$). There are essentially two different fates for the particles: Firstly, a portion of particles reaches the walls of the enclosure and adheres to the surface. Secondly, the buffer gas effectively transports a fraction of the particles toward the outlet, leading to their expulsion from the cavity.

\begin{figure}[h]
	\centering\includegraphics[width=0.9\linewidth]{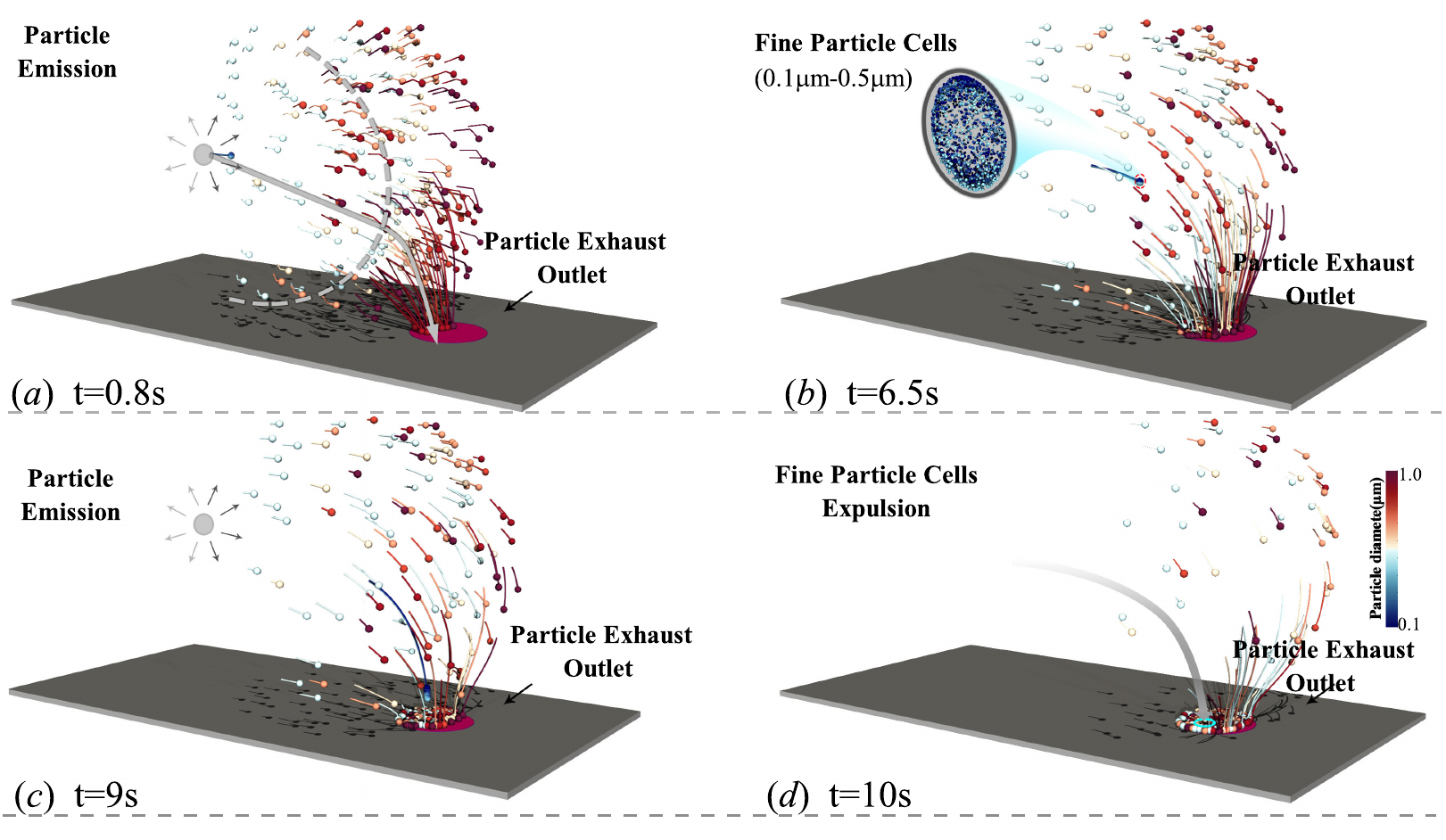}
	\caption{Snapshots of debris particle transport by buffer gas at low inflow velocity $v_{\rm{in}}$=1m/s. Particles are colored with the particle diameter in $\upmu$m, ranging from 0.1 $\upmu$m to 1 $\upmu$m. The displayed particles here are the ones discharged by the buffer gas from the outlet.}
	\label{fig:trajectory_l}
\end{figure}

To identify any preferential particle aggregation regions within the cavity, we analyzed the particle size histograms along the distance on the lower plate. Figure \ref{fig:partical_1}($b$)-($e$) illustrates the resulting statistics, where two prominent aggregation regions are identified. These regions are centered around $y$=0.5 m and $y$=1.5 m, as marked by the dashed lines in the figure. The dashed lines correspond to the longitudinal positions of releasing particles and the outlet position within the cavity. Additionally, it is noteworthy that for low-speed buffer gas, particles with diameters smaller than 0.4 $\upmu$m can be completely expelled from the cavity. However, in the case of high-speed gas flow, small-sized particles ($\le$ 0.4 $\upmu$m) are transported by the strong jet flow and subsequently adhere to the cavity walls due to the impact of the buffer gas. The result suggests that an excessively strong injection flow may lead to an opposite effect.

\begin{figure}[th]
	\centering\includegraphics[width=0.9\linewidth]{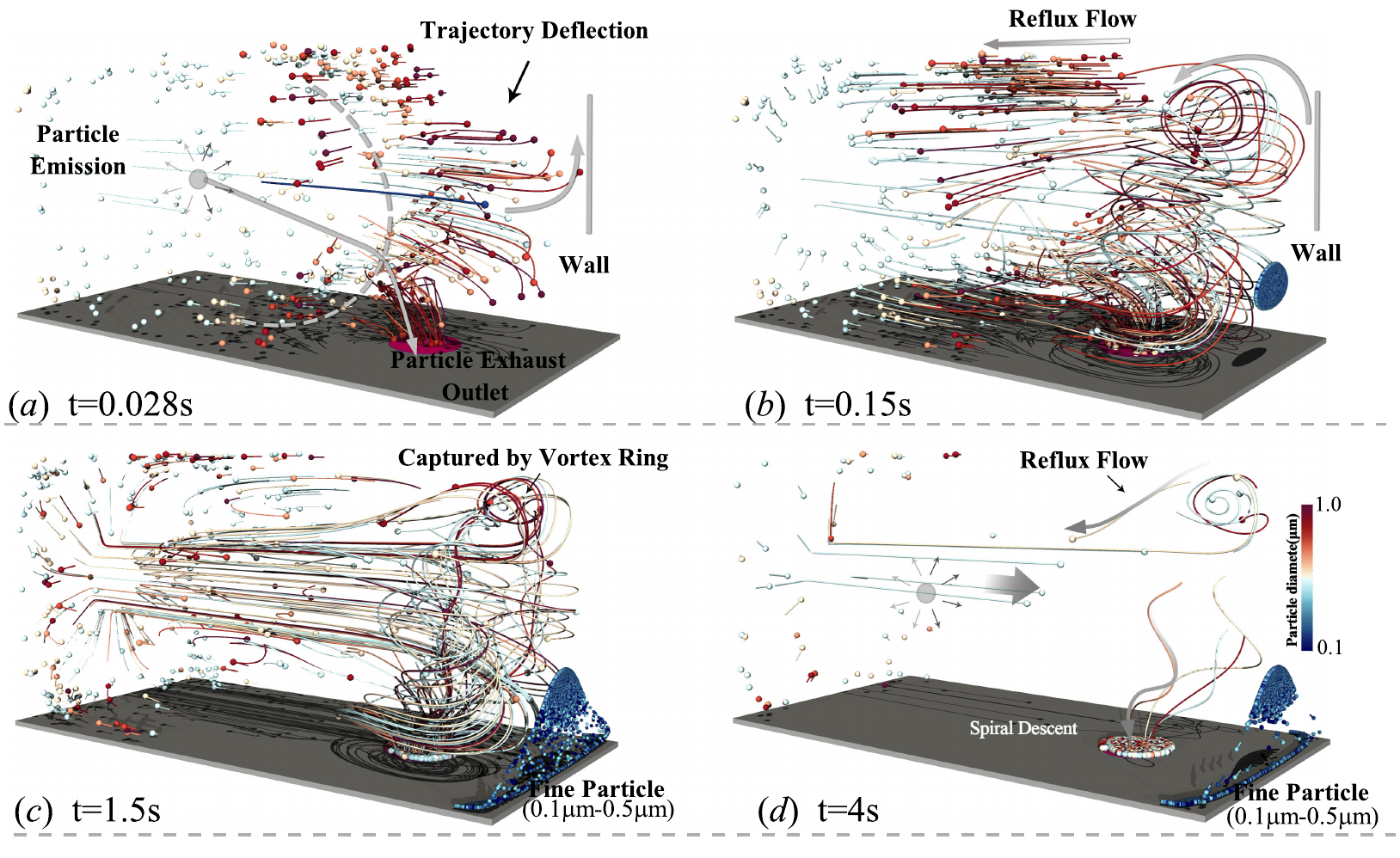}
	\caption{Snapshots of debris particle transport by buffer gas at high inflow velocity $v_{\rm{in}}$=50m/s. Particles are colored with the particle diameter in $\upmu$m, ranging from 0.1 $\upmu$m to 1 $\upmu$m. The displayed particles here are the ones discharged by the buffer gas from the outlet.}
	\label{fig:trajectory_h}
\end{figure}

To closely inspect the movement of debris particles, Fig. \ref{fig:trajectory_l} illustrates the location of the particles at different time instants for low inflow velocity $v_{\rm{in}}$=1m/s, with short-time trajectories drawn to guide the eyes. As depicted in Fig. \ref{fig:trajectory_l}($a$), during the initial injection stage, a cloud of debris particles is generated and expands radially from the particle release position at (0.5$H$, 0.5$H$, 0.5$H$). The dynamics of this radial expansion can be understood by the damping of initial momentum by the drag force, taking into account the rarefaction effect. Smaller particles experience significant deceleration as the inverse of drag goes by a square of the diameter. It results in a highly localized spot transported by the buffer gas (see the magnified region in Fig. \ref{fig:trajectory_l}($b$)). These fine particles become a collective unit without being separated by the buffering flow. Subsequently, these collective particles are gradually carried out of the cavity by the buffer gas (see Figs. \ref{fig:trajectory_l}$c$, $d$). In contrast, larger particles are less influenced by the drag and exhibit a noticeable outward expansion. Additionally, particles located near the outlet show a tendency to move toward the exit under the influence of the buffer gas evacuation.

For a one-to-one comparison, we plot in Fig. \ref{fig:trajectory_h} the particles and their short-time trajectories for a high inflow velocity $v_{\rm{in}}=50$m/s. In contrast to the low inflow velocity case, small-sized particles experience impaction on the wall surfaces under the influence of the strong buffer gas transport. The short-time trajectories also outline the formation of localized vortices, as seen from the circulating motion of particles (see Figs. \ref{fig:trajectory_h}$b$ and $c$). Throughout this intricate process, a fraction of particles is captured by the vortices and re-entrained upstream, while others spiral downward instead of straightly moving and are discharged through the outlet (see Fig. \ref{fig:trajectory_h}$d$). The phenomenon, known as spiral patterns, is commonly observed in nature and widely utilized in industrial settings. For instance, spiral patterns are observed at a downward-facing free surface of a horizontal liquid film \cite{Yoshikawa_prl_2019}. Besides, patterns of stable spirals and chaotic spiral defects are frequently observed in Rayleigh-B\'enard convection \cite{Bodenschatz_prl_1991,Zhou_prl_2007}, along with many other fluid flow systems\cite{Plapp_prl_1998}.

\begin{figure}[h]
	\centering\includegraphics[width=1.0\linewidth]{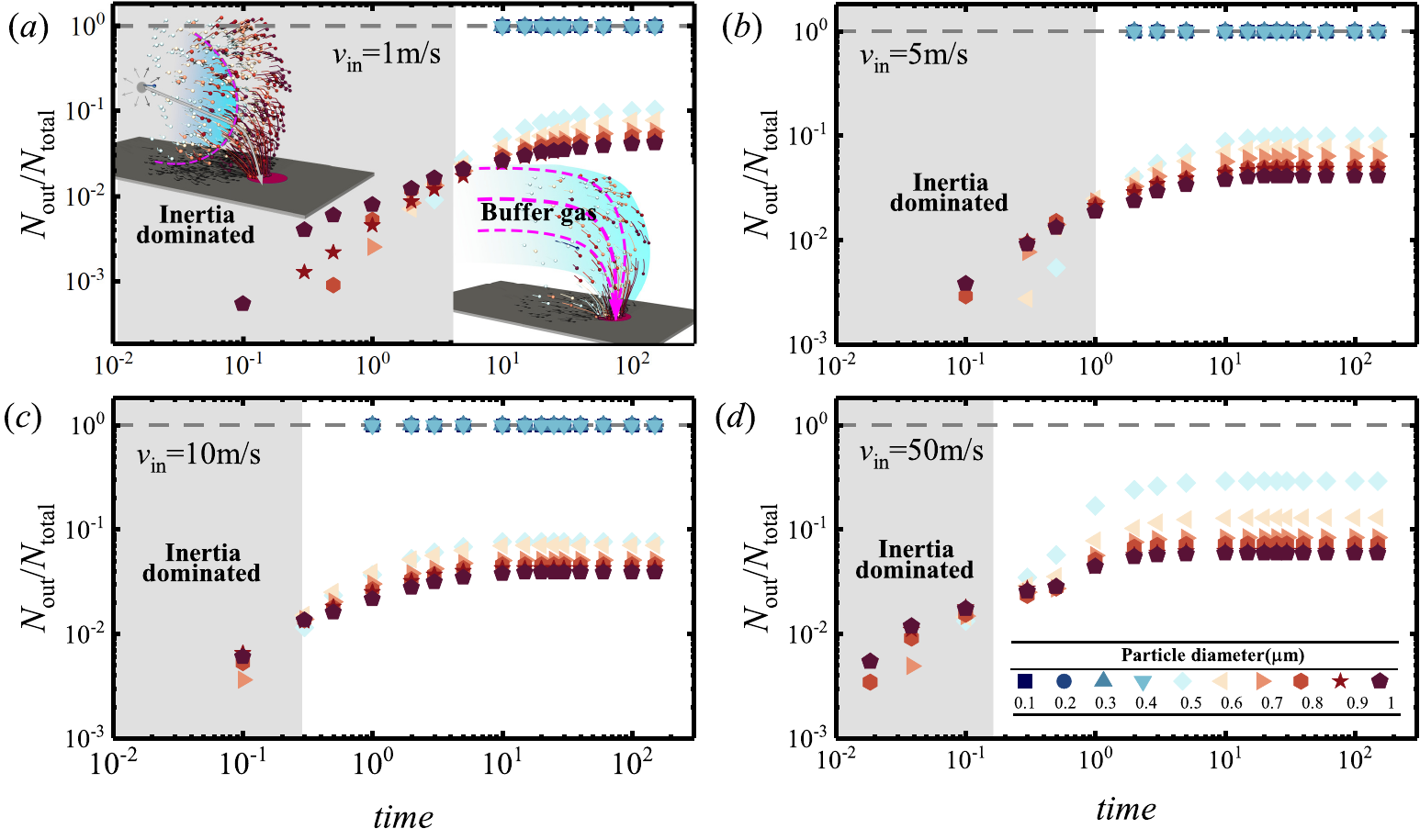}
	\caption{Proportional distribution of particles expelled from the buffer gas discharge cavity at different inflow velocities: ($a$) $v=1$ m/s, ($b$) $v=5$m/s, ($c$) $v=10$m/s, and ($d$) $v=50$m/s.}
	\label{fig:proportional}
\end{figure}

Figure \ref{fig:proportional} illustrates the variation of the number of particles expelled from the cavity over time, categorized by different particle sizes. During the initial stages, there is a relatively higher rate of expulsion for larger particles compared to smaller particles. However, as time progresses, there is a transition, and smaller particles start to be expelled in greater numbers compared to larger particles. Indeed, this phenomenon can be attributed to the dominance of inertial effects in particle motion during the initial short period. Smaller particles experience a significant deceleration within the cavity. In contrast, larger particles are less influenced by drag and exhibit a rapid outward expansion in all directions. Thus, large particles reach the region close to the outlet, allowing those particles to be expelled from the cavity in a shorter period of time. Furthermore, the transition time decreases as the velocity of the buffering gas flow increases (as indicated by the gray markers in Figs. \ref{fig:proportional}$a$-$d$).

\section{Conclusion}\label{Conclusion}
In summary, this study demonstrates the use of buffer gas to promote debris particle transport in EUV source vessel, employing direct numerical simulations (DNS) for active debris clearance with inlet jet flow velocity ranging from $1$m/s to $50$m/s. The rectangular cavity is considered, with buffer gas ($H_2$) entering through a circular opening on the left side and exiting through an outlet on the bottom plate.

Our study reveals the influence of the inflow velocity of buffer gas on the flow morphological transition and its consequences for particle transport within a cavity. For low inflow velocities, the flow field remains steady. However, as the inflow velocity increases, the buffer gas jet induces the formation of clustered corner rolls at the inlet region. The size of these corner rolls increases with the inflow velocity. When the inflow velocity reaches a sufficiently high level, the buffer gas jet penetrates through the entire cavity, impacting the endwall. This impact leads to the formation of a recirculation zone in the upper part of the cavity. Interestingly, the recirculating flow reaches the outlet with the spontaneous formation of spiraling motion. These distinct flow structures also result in a transition in the particle transport mode: for low-speed buffer gas, particles with diameters smaller than 0.4 $\upmu$m can be completely expelled from the cavity. However, in the case of high-speed buffer gas, small-sized particles ($\le 0.4 \upmu$m) are transported by the strong jet flow and subsequently adhere to the cavity walls due to the impact of the buffer gas. The contrasting effects of low-speed and high-speed buffer gas on the expulsion and adherence of particles emphasize the importance of precise control and understanding of flow conditions. The practical implications for optimizing particle transport in confined environments are well emphasized, and the insight in this study hold the potential for informing the development of effective strategies in various applications, especially in the environment of EUV lithography where one relies largely on using buffer gas with high flow rate to flush debris particle with a range of sizes.

\begin{acknowledgments}
This work was sponsored by the Natural Science Foundation of China under Grant Nos. 11988102, 92052201, 91852202, 11825204, 12102246, 12032016, 11972220, and 12372219, the Program of Shanghai Academic Research Leader under Grant No. 19XD1421400, the Shanghai Science and Technology Program under Project Nos. 19JC1412802 and 20ZR1419800. The authors report no conflict of interest.
\end{acknowledgments}

\section*{Data Availability}
The data that support the findings of this study are available from the corresponding authors upon reasonable request.



\bibliography{reference}

\end{document}